%% file: ISWCS_24_JAD.tex
\documentclass[conference,10pt,twoside,twocolumn]{IEEEtran}
\IEEEoverridecommandlockouts
\usepackage{cite}
\usepackage[T1]{fontenc}
\usepackage{graphicx}
\usepackage{amssymb}
\usepackage{amsmath}
\usepackage{amsthm}
\usepackage{subfigure}
\usepackage{booktabs} 
\usepackage{multirow}
\usepackage{microtype}
\usepackage{balance}
\usepackage{xcolor}
\usepackage{dblfloatfix}

\usepackage[hidelinks]{hyperref}

\input{macros/vmr-symbols-vecbold}
\input{macros/standard-macros}

\input{macros/defs}

\begin{document}

\title{Deep-Unfolded Joint Activity and Data Detection for Grant-Free Transmission in Cell-Free Systems}

\author{\IEEEauthorblockN{Gangle~Sun\textsuperscript{1,2}, Wenjin~Wang\textsuperscript{1,2}, Wei~Xu\textsuperscript{1,2}, and Christoph~Studer\textsuperscript{3}}\\
	\textit{\textsuperscript{1}National Mobile Communications Research Laboratory, Southeast University, Nanjing, China}\\
	\textit{\textsuperscript{2}Purple Mountain Laboratories, Nanjing, China}\\
	\textit{\textsuperscript{3}Department of Information Technology and Electrical Engineering, ETH Zurich, Switzerland}\\
	\textit{email: sungangle@seu.edu.cn, wangwj@seu.edu.cn, wxu@seu.edu.cn, and studer@ethz.ch}\\
\thanks{This paper extends our box-constrained FBS algorithm from \cite{sun2023joint_arxiv} through deep unfolding, a momentum strategy, an approximate posterior mean estimator, and a novel soft-output AUD module.}
\thanks{
This work was supported in part by the National Key R\&D Program of China under Grant 2023YFB2904703; in part by the National Natural Science Foundation of China under Grants 62341110, 62371122, and 62022026; in part by the Jiangsu Province Basic Research Project under Grant BK20192002; and in part by the Fundamental Research Funds for the Central Universities under Grants 2242022k30005, 2242022k60002, and 2242023k5003.
The work of Gangle Sun was supported in part by the China Scholarship Council under Grant 202206090074.}
\thanks{The authors would like to thank Sueda Taner and Oscar Casta\~neda for their advice on training deep unfolding networks.}
}

\maketitle

\begin{abstract}
Massive grant-free transmission and cell-free wireless communication systems have emerged as pivotal enablers for massive machine-type communication. This paper proposes a deep-unfolding-based joint activity and data detection (DU-JAD) algorithm for massive grant-free transmission in cell-free systems. We first formulate a joint activity and data detection optimization problem, which we solve approximately using forward-backward splitting (FBS). We then apply deep unfolding to FBS to optimize algorithm parameters using machine learning. In order to improve data detection (DD) performance, reduce algorithm complexity, and enhance active user detection (AUD), we employ a momentum strategy, an approximate posterior mean estimator, and a novel soft-output AUD module, respectively. Simulation results confirm the efficacy of DU-JAD for AUD and DD. 
\end{abstract}

\begin{IEEEkeywords}
Cell-free, deep unfolding, joint activity and data detection, 
massive machine-type communication.
\end{IEEEkeywords}
\section{Introduction}
\label{sec:intro}
Massive machine-type communication (mMTC) is a core component of fifth-generation (5G) wireless communication systems. mMTC is characterized by intermittent data transmissions from user equipments (UEs) to an infrastructure basestation (BS). 
Massive grant-free transmission techniques have proven to be effective for mMTC scenarios, as they alleviate signaling overhead, network congestion, and transmission latency by facilitating direct signal transmission from active UEs over shared resource elements, avoiding the need for intricate scheduling mechanisms~\cite{sun2023joint_arxiv, Ke2021massive, Sun2022massive,sun2021OFDMA}.

To improve coverage for UEs in mMTC scenarios, cell-free communication has emerged as a powerful solution~\cite{Mishra2022rate, xu2023toward}.
Cell-free communication ameliorates inter-cell interference and boosts spectral efficiency through joint processing of received signals from distributed access points (APs), which interface with a central processing unit (CPU)~\cite{Ke2021massive, song2022joint, xu2023edge}.
For massive grant-free transmission in cell-free systems, crucial tasks encompass active user detection (AUD) and data detection (DD) at the CPU side.

\newcounter{TempEqCnt}
\setcounter{TempEqCnt}{\value{equation}}
\setcounter{equation}{6}
\begin{figure*}
	\vspace{-1ex}
	\begin{equation}
		\begin{aligned}
		\mathcal{P}_1:\;\Big\{ {\hat{\mathbf{H}},{{\hat {\mathbf{X}}}_{\text{D}}}} \Big\} =\mathop {\arg\min }_{\scriptstyle{\mathbf{H}\in\mathbb{C}^{MP\times N}}\hfill\atop
				{	\scriptstyle{{\mathbf{X}}_{\text{D}}\in\mathcal{B}}\hfill}} \frac{1}{2} \left\|\mathbf{Y}-\mathbf{H}\left[\mathbf{X}_{\text{P}},\mathbf{X}_{\text{D}}\right]\right\|_F^2 
			+ \mu_h\sum_{n=1}^N\sum_{p=1}^P\left\|\mathbf{h}_{n,p}\right\|_F+ \mu_x\sum_{n=1}^N\left\|\mathbf{x}_{\text{D},n}\right\|_F.
		\end{aligned}
		\label{P1}
	\end{equation}
	\vspace{-3ex}
	\hrulefill
\end{figure*}
\setcounter{equation}{\value{TempEqCnt}}

\subsection{Contributions}
We propose a deep-unfolding-based joint activity and data detection (DU-JAD) algorithm for massive grant-free transmission in cell-free systems.
First, we adapt the box-constrained forward-backward splitting (FBS) algorithm from~\cite{sun2023joint_arxiv} to approximately solve a joint activity and data detection (JAD) problem, followed by deep-unfolding (DU) its iterations, thereby enabling automated parameter tuning using machine learning. 
Each unfolded FBS iteration utilizes separate step sizes and a momentum strategy to improve JAD performance. 
Second, we include an approximate posterior mean estimator (PME) to improve DD.
Finally, we propose a new soft-output AUD module that improves user activity detection.

\subsection{Relevant Prior Art}

Recent research has concentrated on massive grant-free transmission in cell-free communication systems.
References \cite{Ganesan2021algorithm} and \cite{shao2020covariance} achieved effective AUD by capitalizing on prominent APs and exchanging activity information between neighboring APs, respectively.
Reference \cite{Li2023Asynchronous} focused on the AUD task in asynchronous transmission scenarios. 
Reference \cite{Diao2023Scalable} proposed the deep learning-based AUD method to achieve near-real-time transmission.
Both \cite{Jiang2023EM} and \cite{wang2022two} relied on separate  AUD and channel estimation (CE) tasks with approximate message passing.
Moreover, \cite{Ke2021massive}, \cite{guo2022joint}, and \cite{Johnston2022model} explored joint AUD and CE, where \cite{guo2022joint} proposed individual AP signal processing, \cite{Ke2021massive} addressed quantization artifacts, and \cite{Johnston2022model} utilized channel sparsity in millimeter-wave systems.
Reference~\cite{Di2022adaptive} delved into JAD executed locally at each AP, followed by the culminating joint detection at the CPU based on an adaptive AP selection method.
Reference~\cite{Iimori2021grant} performed joint AUD, CE, and DD through bilinear Gaussian belief propagation, and our previous work in~\cite{sun2023joint_arxiv} proposed a box-constrained FBS algorithm for these purposes. 
DU techniques~\cite{Balatsoukas2019deep} have also gained prominence in this space~\cite{Johnston2022model}, which combines backpropagation with stochastic gradient descent to optimize algorithm parameters.

In contrast, we focus on JAD for massive grant-free transmission in cell-free systems and propose DU-JAD by utilizing deep unfolding to optimize all algorithm parameters. 
Furthermore, we include a momentum strategy and an approximate PME, followed by a novel soft-output AUD module that processes the estimated channel and data symbols. 

\subsection{Notation}

Matrices are denoted by uppercase boldface letters, column vectors by lowercase boldface letters, and sets by uppercase calligraphic letters. 
The entry at the $m$th row and $n$th column of matrix~$\mathbf{A}$ is $\mathbf{A}(m,n)$ ;the $m$th row vector of matrix~$\mathbf{A}$
is $\mathbf{A}(m,:)$.
The superscripts $(\cdot)^*$, $(\cdot)^T$, and $(\cdot)^H$ denote conjugate, transpose, and conjugate transpose, respectively;  \(\mathbf{I}_N\) is the \(N \times N\) identity matrix.
A diagonal matrix with entries $\{x_1,\ldots,x_N\}$ on the main diagonal is $\operatorname{diag}\{x_1,\ldots,x_N\}$.
The indicator function $\mathbb{I}\{\cdot\}$ is one if its condition is true and zero otherwise.
The $\ell_1$ norm and Frobenius norm are $\|\cdot\|_1$ and $\|\cdot\|_F$, respectively.
The cardinality of the set $\mathcal{Q}$ is $|\mathcal{Q}|$. 

\section{System Model and JAD}
\label{sec:system}
We consider a cell-free wireless communication system with frequency-flat and block-fading channels, in which $P$ distributed APs equipped with $M$ antennas each serve $N$ single-antenna UEs.
The UEs show sporadic activity, and active UEs occupy~$R$ shared resources to transmit their uplink signals.

\subsection{System Model}
Following our previous work in~\cite{sun2023joint_arxiv}, the received signals of all APs, $\mathbf{Y}\in\mathbb{C}^{MP\times R}$, can be expressed as follows:
\begin{equation}
	\begin{aligned}
		 \mathbf{Y} = \sum_{n=1}^{N}\xi_n \mathbf{h}_{n}\bar{\mathbf{x}}_n^T + \mathbf{N}
		= \mathbf{H}\mathbf{X}+ \mathbf{N}.
	\end{aligned}
	\label{pilot_data}
\end{equation}
Here, $\xi_n\in\{0,1\}$ is the activity indicator of the $n$th UE with $\xi_n=1$ indicating  active state and $0$ otherwise.
The channel vector between APs and the $n$th UE is
$\mathbf{h}_{n}=[\mathbf{h}_{n,1}^T,\mathbf{h}_{n,2}^T,\ldots,\mathbf{h}_{n,P}^T]^T\in\mathbb{C}^{MP}$ with $\mathbf{h}_{n,p}\in\mathbb{C}^{M}$ representing the channel vector between the $p$th AP and $n$th UE.
The signal vector of the $n$th UE is
$\bar{\mathbf{x}}_{n}=[\bar{\mathbf{x}}_{\text{P},n}^T,\bar{\mathbf{x}}_{\text{D},n}^T]^T\in\mathbb{C}^{R}$ with $\bar{\mathbf{x}}_{\text{P},n}\in\mathbb{C}^{R_{\text{P}}}$ and $\bar{\mathbf{x}}_{\text{D},n}\in\mathcal{Q}^{R_{\text{D}}}$ being the pilot and data vectors, respectively. 
In what follows, we consider quadrature phase shift keying (QPSK) data transmission with constellation set $\mathcal{Q}\triangleq\{x:\,x=\pm B \pm j B\}$ with $B>0$.
The noise matrix is 
$\mathbf{N}\in\mathbb{C}^{MP\times R}$ and contains i.i.d. circularly-symmetric complex Gaussian entries with variance $N_0=1$. 
For simplicity, we define an equivalent channel matrix $\mathbf{H}\triangleq\left[\xi_1\mathbf{h}_{1},\xi_2\mathbf{h}_{2},\ldots,\xi_N\mathbf{h}_{N}\right]\in\mathbb{C}^{MP\times N}$ and signal matrix $\mathbf{X}\triangleq\left[\mathbf{X}_{\text{P}},\mathbf{X}_{\text{D}}\right]\in\mathbb{C}^{N\times R}$, where $\mathbf{X}_{\text{P}} = [\bar{\mathbf{x}}_{\text{P},1},\bar{\mathbf{x}}_{\text{P},2},\ldots,\bar{\mathbf{x}}_{\text{P},N}]^T\in\mathbb{C}^{N\times R_{\text{P}}}$ is the non-sparse pilot matrix to ensure full use of known information and $\mathbf{X}_{\text{D}} = [\mathbf{x}_{\text{D},1},\mathbf{x}_{\text{D},2},\ldots,\mathbf{x}_{\text{D},N}]^T\in\bar{\mathcal{Q}}$ is the row-sparse data~matrix with $\mathbf{x}_{\text{D},n}\triangleq\xi_n\bar{\mathbf{x}}_{\text{D},n}$ and $\bar{\mathcal{Q}}\triangleq\{\mathbf{X}\in\mathbb{C}^{N\times R_{\text{D}}}:\,\mathbf{X}(n,:)^T \in \{\mathcal{Q}^{R_{\text{D}}}, \mathbf{0}\}, \,\forall n\}$. 
We emphasize that the sporadic user activity and inherent channel sparsity among APs render the matrix \( \mathbf{H} \) block-sparse.

\subsection{Problem Formulation}
While our primary objective is JAD, a precise estimation of the equivalent channel matrix $\mathbf{H}$ improves  JAD performance. Therefore, we formulate the JAD optimization problem for massive grant-free transmission in cell-free systems as in~\cite{sun2023joint_arxiv}
\begin{align}
\{ \hat{\mathbf{H}},{{\hat {\mathbf{X}}}_{\text{D}}} \} =\mathop {\arg \max }\limits_{\scriptstyle{\mathbf{H}\in\mathbb{C}^{MP\times N}}\hfill\atop
		{	\scriptstyle{{\mathbf{X}}_{\text{D}}\in\bar{\mathcal{Q}}}\hfill}} P\!\left(\mathbf{Y}|\mathbf{H},\mathbf{X}_{\text{D}}\right)\!P\!\left(\mathbf{H}\right)\!P\!\left(\mathbf{X}_{\text{D}}\right),
	\label{joint_problem}
\end{align}
with 
\begin{equation}
P\!\left(\mathbf{Y}|\mathbf{H},\mathbf{X}_{\text{D}}\right) \propto\exp\left(-\frac{\left\|\mathbf{Y}-\mathbf{H}\left[\mathbf{X}_{\text{P}},\mathbf{X}_{\text{D}}\right]\right\|_F^2}{N_0}\right)\!. 
\end{equation}
As in \cite{sun2023joint_arxiv} and \cite{song2022joint}, we leverage the following complex-valued block-Laplace model and complex-valued Laplace model to take the sparsity of $\mathbf{H}$ and $\mathbf{X}_{\text{D}}$ into account:
\begin{align}
    &P\!\left(\mathbf{H}\right) \propto \prod_{n=1}^{N}\prod_{p=1}^{P}\exp\left(-2\mu_h\left\|\mathbf{h}_{n,p}\right\|_F\right),\\
    &P\!\left(\mathbf{X}_{\text{D}}\right) \propto \prod_{n=1}^{N}\exp\left(-2\mu_x\left\|\mathbf{x}_{\text{D},n}\right\|_F\right),
\end{align}
where $\mu_h$ and $\mu_x$ are parameters that determine the sparsity of $\mathbf{H}$ and~$\mathbf{X}_{\text{D}}$, respectively. 
The discrete nature $\mathbf{X}_{\text{D}}$ due to the constellation $\bar{\mathcal{Q}}$ renders $\mathcal{P}_1$ a discrete-valued optimization problem and a na\"ive exhaustive search is simply infeasible. To arrive at a computationally manageable optimization problem, we relax the set $\bar{\mathcal{Q}}$ to its convex hull~\cite{song2022joint}:
\begin{equation}
    \mathcal{B}=\left\{\sum_{i=1}^{|\bar{\mathcal{Q}|}}\delta_i \mathbf{X}_i:\,\mathbf{X}_i\in\bar{\mathcal{Q}},\delta_i\ge 0,\forall i;\,\sum_{i=1}^{|\bar{\mathcal{Q}|}}\delta_i=1\right\}\!.
\end{equation}

By inserting the above probability density functions into~(\ref{joint_problem}) and by relaxing the constellation set to its convex hull, we obtain the optimization problem $\mathcal{P}_1$ in (\ref{P1}), which we solve approximately with FBS~\cite{goldstein2014field}.

\subsection{FBS-based JAD Algorithm}

With the definition $\mathbf{S}\triangleq[\mathbf{H}^H, \mathbf{X}_{\text{D}}]^H\in\mathbb{C}^{(MP+R_{\text{D}})\times N}$, the FBS-based JAD algorithm starts by splitting the objective function in $\mathcal{P}_1$ into 
\setcounter{equation}{7}
\begin{align}
    &f(\mathbf{S})= \frac{1}{2}\left\|\mathbf{Y}-\mathbf{H}\left[\mathbf{X}_{\text{P}},\mathbf{X}_{\text{D}}\right]\right\|_F^2,\\
    &g(\mathbf{S})= \mu_h\sum_{n=1}^N\sum_{p=1}^P\left\|\mathbf{h}_{n,p}\right\|_F + \mu_x\sum_{n=1}^N\left\|\mathbf{x}_{\text{D},n}\right\|_F+ \mathcal{X}(\mathbf{X}_{\text{D}}),
\end{align}
where $\mathcal{X}\!\left(\mathbf{X}_{\text{D}}\right)\!=+\infty\,\mathbb{I}\{\exists \mathbf{X}_{\text{D}} \notin \mathcal{B}\}$ enforces the data to be within the convex set $\mathcal{B}$.
FBS alternates between a gradient-descent step in the smooth function $f(\mathbf{S})$ (the forward step) and performing a proximal operation to find a solution near the minimizer of $g(\mathbf{S})$ (the backward step), where iterations are carried out until a convergence criterion is met. The forward and backward steps are as follows:

{\textbf{\textit{1) Forward Step:}}} The forward step is given by 
\begin{equation}
\hat{\mathbf{S}}^{k}=\mathbf{S}^k-\tau^k \nabla f \!\left(\mathbf{S}^k\right)\!,
\end{equation}
where the superscript $k$ indicates the $k$th iteration, $\hat{\mathbf{S}}^{k} = [(\hat{\mathbf{H}}^{k})^H,\hat{\mathbf{X}}_{\text{D}}^{k}]^H$, $\tau^k$ is the step size~\cite{goldstein2014field}, and the gradient of $f \!\left(\mathbf{S}\right)\!$ with respect to $\mathbf{S}$ is given by 
\begin{equation}
    \nabla f (\mathbf{S}) =-\left[\mathbf{X}^*\!\left(\mathbf{Y}^T-\mathbf{X}^T\mathbf{H}^T\right)\! ,\mathbf{H}^T\!\left(\mathbf{Y}_{\text{D}}-\mathbf{H}\mathbf{X}_{\text{D}}\right)\!^* \right]^T.
\end{equation}

{\textbf{\textit{2) Backward Step:}}}
The proximal operator for $\mathbf{H}$ is 
\begin{equation}
    \mathbf{h}_{n,p}^{k+1} = \frac{\max\left\{\left\|\hat{\mathbf{h}}_{n,p}^{k}\right\|_F-\tau^k\mu_h,0\right\}\hat{\mathbf{h}}_{n,p}^{k}}{\left\|\hat{\mathbf{h}}_{n,p}^{k}\right\|_F},
\end{equation}
and the proximal operation for $\mathbf{X}_{\text{D}}$ involves derivations using the Karush-Kuhn-Tucker conditions; see \cite{sun2023joint_arxiv} for the details.

\section{DU-JAD: Deep-Unfolding-based joint activity and data detection}
\label{sec:DU-JAD}

\begin{figure}
	\centering
	\includegraphics[width=0.48\textwidth]{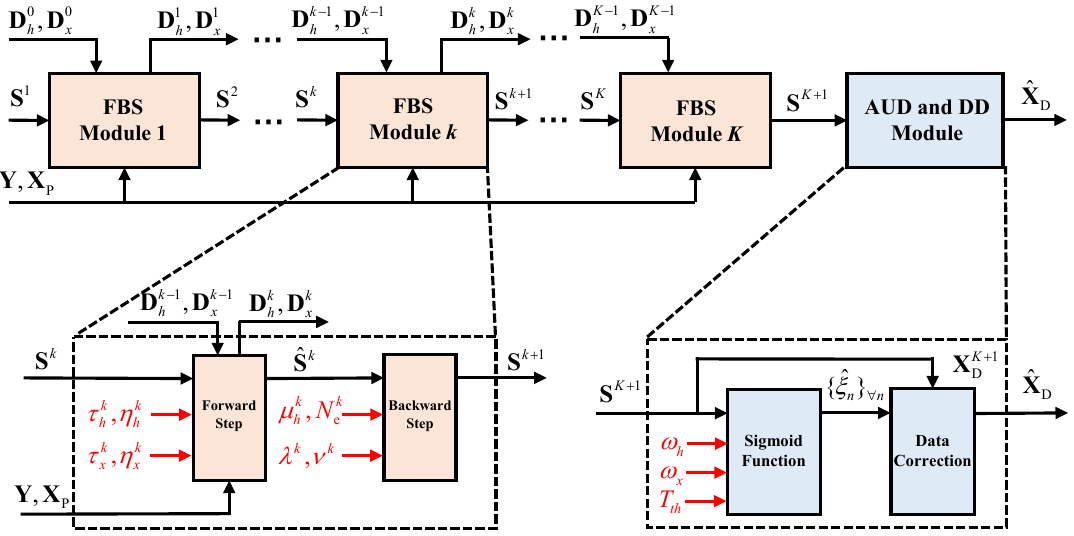}
	\caption{Architecture of the DU-JAD algorithm.}
	\label{DU-JAD}
\end{figure}

Manually tuning the parameters of FBS-based JAD  is, in general, tedious. Thus, we apply DU~\cite{Balatsoukas2019deep}, resulting in DU-JAD, which enables automated parameter tuning.
DU transforms the iterative procedure of the FBS-based JAD algorithm into a sequential computation process traversing multiple modules with identical structure. The general architecture of DU-JAD is depicted in Fig.~\ref{DU-JAD}.

\subsection{FBS Module}
\phantom{gangle} \\[-0.9cm]

{\textbf{\textit{1) Forward Step:}}} In the forward step of the FBS-based JAD algorithm, the step size $\tau^k$ is used when computing $\hat{\mathbf{H}}^{k}$ and $\hat{\mathbf{X}}_{\text{D}}^{k}$. Considering the vastly different value ranges of $\mathbf{H}$ and $\mathbf{X}_{\text{D}}$, we introduce two separate step sizes $\tau_h^k$ and $\tau_x^k$ in DU-JAD for the computation of $\hat{\mathbf{H}}^{k}$ and $\hat{\mathbf{X}}_{\text{D}}^{k}$, respectively. 
To further accelerate convergence, we include a momentum term with weights $\eta_h^k$ and $\eta_x^k$. Consequently, the forward step in the $k$th FBS module of DU-JAD is given by
	\begin{align}
		&\hat{\mathbf{H}}^{k}=\mathbf{H}^k+\tau_h^k \!\left(\mathbf{Y}-\mathbf{H}^k\mathbf{X}^k\right)\!\left(\mathbf{X}^k\right)^H + \eta_h^k \mathbf{D}_h^{k-1},\\
		&\hat{\mathbf{X}}_{\text{D}}^{k}=\mathbf{X}_{\text{D}}^k+\tau_x^k \!\left(\mathbf{Y}_{\text{D}}-\mathbf{H}^k\mathbf{X}_{\text{D}}^k\right)\!^H\mathbf{H}^k + \eta_x^k \mathbf{D}_x^{k-1},
	\end{align}
where 
\begin{align}
    &\mathbf{D}_h^k=\tau_h^{k} \!\left(\mathbf{Y}-\mathbf{H}^{k}\mathbf{X}^{k}\right)\!\left(\mathbf{X}^{k}\right)^H + \eta_h^{k} \mathbf{D}_h^{k-1},\\
    &\mathbf{D}_x^k=\tau_x^{k} \!\left(\mathbf{Y}_{\text{D}}-\mathbf{H}^{k}\mathbf{X}_{\text{D}}^{k}\right)\!^H\mathbf{H}^{k} + \eta_x^{k} \mathbf{D}_x^{k-1},
\end{align}
with $\mathbf{D}_h^{0}=\mathbf{0}$
and $\mathbf{D}_x^{0}=~\mathbf{0}$.

{\textbf{\textit{2) Backward Step:}}} In the backward step of the FBS-based JAD algorithm corresponding to $\mathbf{H}$, the parameter $\mu_h$ is fixed across iterations. To increase flexibility in the FBS-based JAD algorithm, we untie the parameter $\mu_h$ as $\{\mu_h^k\}_{\forall k}$ over distinct FBS modules.
In the backward step of the FBS-based JAD algorithm corresponding to $\mathbf{X}_{\text{D}}$, the estimated data matrix may not reside within the set $\bar{\mathcal{Q}}$ due to relaxation of the discrete set $\bar{\mathcal{Q}}$ to its convex hull $\mathcal{B}$. To take into account the discrete nature of the constellation $\bar{\mathcal{Q}}$, we propose to use the posterior mean estimation method for DD instead of introducing the penalty term into the objective function as done in \cite{sun2023joint_arxiv}.

We start by rewriting the $n$th UE's estimated data vector $\hat{\mathbf{x}}_{\text{D},n}^{k}$ in the $k$th iteration as $
	\hat{\mathbf{x}}_{\text{D},n}^{k} = \mathbf{x}_{\text{D},n} + \mathbf{e}^{k}$,
where $\mathbf{x}_{\text{D},n}$ is the unknown data vector and $\mathbf{e}^{k}$ is the estimation error in iteration $k$. By assuming  $\mathbf{e}^{k}$ follows a complex Gaussian distribution with mean vector $\mathbf{0}$ and covariance matrix $N_{\text{e}}^k\mathbf{I}_{R_{\text{D}}}$, the PME of $\mathbf{x}_{\text{D},n}$ is given by
\begin{equation}
\mathbf{x}_{\text{D},n}^{k}=\frac{\sum_{\mathbf{x}\in\mathcal{Q}^{R_{\text{D}}}}P_{\text{PME}}^{n,k}(\mathbf{x})\,{P_a}/{|\mathcal{Q}^{R_{\text{D}}}|}\,\mathbf{x}}{\sum_{\mathbf{x}\in\mathcal{Q}^{R_{\text{D}}}}P_{\text{PME}}^{n,k}(\mathbf{x})\,{P_a}/{|\mathcal{Q}^{R_{\text{D}}}}| + P_{\text{PME}}^{n,k}(\mathbf{0})(1-P_a)},
\end{equation}
where $P_a$ represents the user activity probability following $P(\mathbf{X}_{\text{D}})$, and the function $P_{\text{PME}}^{n,k}(\mathbf{x})$ is defined as
\begin{equation}
    P_{\text{PME}}^{n,k}(\mathbf{x})\triangleq \exp\left(-\frac{\left\|\mathbf{x}-\hat{\mathbf{x}}_{\text{D},n}^{k}\right\|_F^2}{N_{\text{e}}^k}\right).
\end{equation}
Evaluating the PME expression of $\mathbf{x}_{\text{D},n}$  incurs high complexity and prone to numerical stability issues. Consequently, we employ an approximate PME as $ \mathbf{x}_{\text{D},n}^{k}=\alpha_n^k\,\check{\mathbf{x}}_{\text{D},n}^{k}$. Here, the scaling factor $\alpha_n^k\in[0,1]$ can be expressed as
\begin{equation}
	 \alpha_n^k=\min\left\{\frac{1}{\left\|\hat{\mathbf{x}}_{\text{D},n}^{k}\right\|_1}\max\{\lambda^k\|\hat{\mathbf{x}}_{\text{D},n}^{k}\|_1-\nu^k,0\},1\right\},
\end{equation}
where $\lambda^k$ and $\nu^k$ are trainable parameters. Besides, the vector $\check{\mathbf{x}}_{\text{D},n}^{k}$ is defined as
\begin{equation}
\check{\mathbf{x}}_{\text{D},n}^{k}\triangleq \frac{\sum_{\mathbf{x}\in\mathcal{Q}^{R_{\text{D}}}}P_{\text{PME}}^{n,k}(\mathbf{x})\,\mathbf{x}}{\sum_{\mathbf{x}\in\mathcal{Q}^{R_{\text{D}}}}P_{\text{PME}}^{n,k}(\mathbf{x})},
\end{equation}
with the estimation error variance  $N_{\text{e}}^k$ to be trained by~DU.

\subsection{AUD and DD Modules}

To improve AUD, we propose a soft-output module that delivers the probability of each user being active instead of carrying out hard-output decisions. To this end, we propose to use a sigmoid  within the AUD module, which leverages the estimated $\mathbf{S}^{K+1}$ to output the likelihood of the user activity status as 
\begin{equation}
    L_n = \frac{1}{1+1/\exp\left(\omega_h\left\|\mathbf{h}_n^{K+1}\right\|_F^2+\omega_x\left\|\mathbf{x}_{\text{D},n}^{K+1}\right\|_F^2-T_{\text{th}}\right)},
\end{equation}
with trainable parameters $\omega_h$, $\omega_x$ and $T_{\text{th}}$.
From the calculated activity probability, the user's active state can be detected by comparing it to a threshold $ \bar{L}\in[0,1]$, i.e.,
\begin{equation}
    \hat{\xi}_n = \mathbb{I}\left\{L_n>\bar{L}\right\}.
\end{equation}

As for the DD module, we detect the data matrix as 
\begin{equation}
    \hat{\mathbf{X}}_{\text{D}}=\operatorname{diag}\{\hat{\xi}_1,\ldots,\hat{\xi}_N\}\tilde{\mathbf{X}}_{\text{D}},
\end{equation}
where $ \tilde{\mathbf{X}}_{\text{D}}=\arg\min_{\mathbf{X}\in\mathcal{Q}^{N\times R_{\text{D}}}}\|\mathbf{X}-\mathbf{X}_{\text{D}}^{K+1}\|_F^2$.

\subsection{Training Procedure}

In DU-JAD, the FBS modules and AUD module undergo distinct training processes. For the FBS iterations (equal to the number of modules), we adopt the squared Frobenius norm of the discrepancy between $\mathbf{X}_{\text{D}}$ and  $\mathbf{X}_{\text{D}}^{K+1}$ as the loss function for parameter training across $K$ FBS modules, i.e., 
\begin{equation}
\operatorname{Loss}_{\text{FBS}}=\left\|\mathbf{X}_{\text{D}}^{K+1}-\mathbf{X}_{\text{D}}\right\|_F^2.
\end{equation}
For the AUD module's parameter training, we employ the empirical binary cross-entropy (BCE) as the loss: 
\begin{equation}
    \operatorname{Loss}_{\text{AUD}}=-\sum_{n=1}^N \xi_n\log(L_n) + (1-\xi_n)\log(1-L_n).
\end{equation}

\section{Simulation Results}
\label{sec:simu}

\subsection{System Setup, Performance Metrics, and Baselines}

As in \cite{sun2023joint_arxiv}, we consider a cell-free communication system with $N = 400$ UEs at a height of 1.65\,m height and $P = 20$ to $100$ APs at a height of 15\,m, each with $M = 4$ antennas, all uniformly distributed in a 500\,m $\times$ 500\,m area. The UE activity $ \{\xi_n\}_{\forall n}$ follows an i.i.d. Bernoulli distribution with $ P_a = 0.2 $. Active UEs transmit $ R_{\text{P}} = 50 $ pilots, based on a complex equiangular tight frame, and $ R_{\text{D}} = 200 $ QPSK signals, i.e., $ B = \sqrt{0.5} $, over a 20\,MHz bandwidth channel at 1.9\,GHz. The UEs have 0.1\,W transmit power and adhere to a 12\,dB power control range. The system works under 8\,dB shadow fading variance, 9\,dB noise figure, and a 290\,K noise temperature. Refer to \cite{sun2023joint_arxiv} for more details.

To evaluate the performance of the proposed DU-JAD,  we consider the user detection error rate (\textit{UDER}) and the average symbol error rate (\textit{ASER}):
\begin{align}
    &\textit{UDER}= \frac{1}{N}\sum_{n=1}^N\left|\xi_n-\hat{\xi}_n\right|,\\
    &\textit{ASER}= \frac{1}{R_{\text{D}}N_a}\sum_{n=1}^N\sum_{r=1}^{R_{\text{D}}}\xi_n\mathbb{I}\left\{\mathbf{X}_\text{D}(n,r)\ne \tilde{\mathbf{X}}_\text{D}(n,r)\right\}.
\end{align}
To demonstrate the efficacy of our algorithm, we compare it to the following baselines: ``{joint AUD-CE via \cite{goldstein2014field}, then DD},'' ``{joint AUD-CE via \cite{chen2018sparse}, then DD},'' ``{joint AUD-CE-DD via \cite{song2022joint}},'' and ``{box-constrained FBS algorithm \cite{sun2023joint_arxiv}},'' which are abbreviated as ``{Baseline 1}'' through ``{Baseline 4}.'' 
In {Baseline 1\,-\,4}, active UEs are identified by evaluating whether the estimated channel energy, $\|\hat{\mathbf{h}}_n^{K+1}\|_F^2$, exceeds the threshold $T_{\text{AUD}}$, setting $\hat{\xi}_n$ to $1$ if true, and $0$ otherwise.
Given the estimated channel matrix and active UEs via {Baseline 1} and {2}, data detection first performs zero-forcing equalization followed by mapping the result to the nearest QPSK symbol. 
To accelerate convergence, we take the result of the {Baseline 1} as the starting point for {Baseline 3}, {Baseline 4} and our DU-JAD algorithm.  
In DU-JAD, the number of FBS modules is fixed to $ K = 10$, while {Baseline 1\,-\,4} use $200$ iterations with a stopping tolerance of $10^{-3}$ for FBS. For comparison with DU-JAD, we also provide the results of {Baseline 2~to~4} with only $10$ iterations.
In the results shown next, we perform~$5000$ Monte--Carlo trials.

\subsection{Simulation Results}
\begin{figure}[t]
	\centering
	\includegraphics[width=0.48\textwidth]{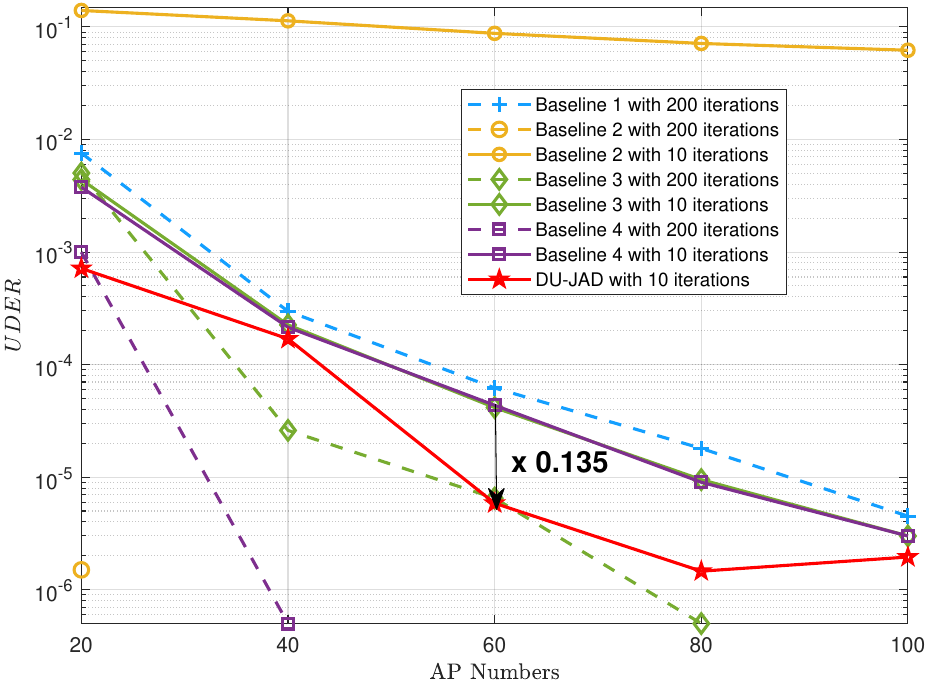}
	\caption{Active user detection performance.}
	\label{AUD}
\end{figure}

\begin{figure}[t]
	\centering
	\includegraphics[width=0.48\textwidth]{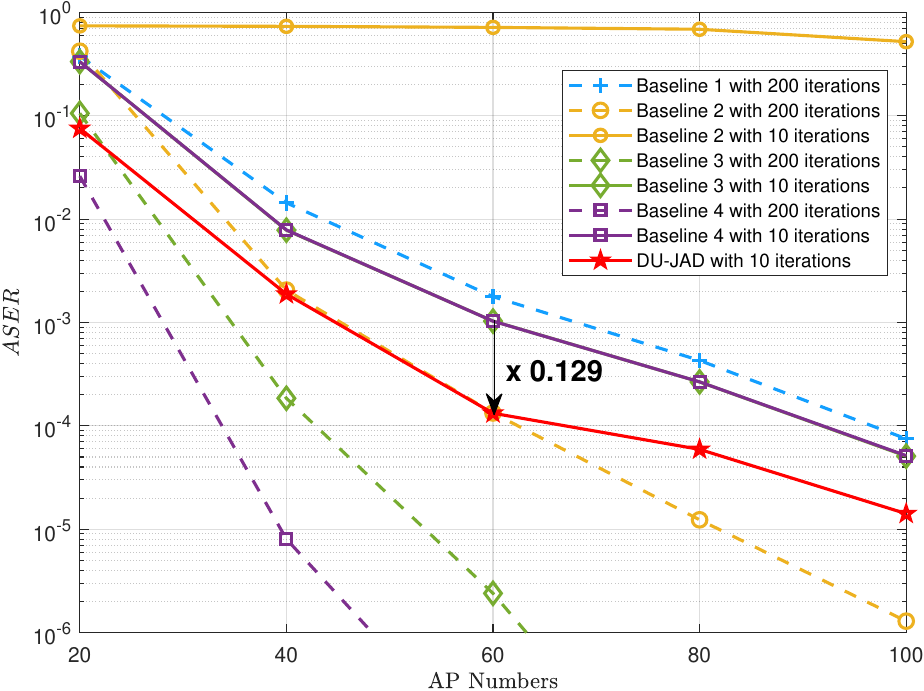}
	\caption{Data detection performance.}
	\label{DD}
\end{figure}

In Fig. \ref{AUD}, we compare the AUD performance for all considered algorithms. 
While {Baseline 2} with $200$ iterations shows the best AUD performance within these methods, its performance drops sharply and becomes the worst when reducing the number of iterations to a more realistic number of $10$ iterations. Similarly, {Baselines 3} and {4}  with $200$ iterations yield excellent AUD performance, which significantly deteriorates when running $10$ iterations. 
Compared with these baselines at $10$ iterations, the proposed DU-JAD method improves AUD substantially. 
Notably, the \textit{UDER} of DU-JAD is only $0.135$ of {Baseline 3} and {4} at $P=60$. In addition, it is worth highlighting that the AUD performance of DU-JAD exhibits variability with changes in the number of APs, e.g., its \textit{UDER} at \(P=100\) is inferior to that at \(P=80\). This variation can be attributed to the AUD module's processing of both estimated channel and data symbols. Since the FBS modules do not include the channel estimation accuracy as an optimization objective, the resulting channel estimates may be inconsistent, which affects the efficacy of the soft-output AUD module.

In Fig. \ref{DD}, we compare the DD performance for all of these methods. While {Baseline 3} and {4} demonstrate better DD performance at $200$ iterations, their effectiveness deteriorates if limited to a more practical number of only $10$ iterations. At the same time, the performance of Baseline 2 collapses completely when running $10$ iterations. Compared to these baselines with $10$ iterations, the proposed DU-JAD method significantly improves DD performance. Specifically, the \textit{ASER} of DU-JAD is only $0.129$ of {Baseline 3} and {4} at $P=60$.

\section{Conclusions}
\label{sec:conclusion}

We have proposed a novel deep-unfolding-based joint activity and data detection (DU-JAD) algorithm for massive grant-free transmission in cell-free wireless communication systems. 
When running only $10$ algorithm iterations, our proposed DU-JAD method (often significantly) outperforms existing baseline methods in terms of active user detection and data detection performance. This is a result of utilizing deep unfolding, incorporating a momentum strategy, deploying an approximate posterior mean estimator, and using a novel soft-output user activity module.

\balance

\bibliographystyle{IEEEtran}
\bibliography{./bib/Refabrv,./bib/IEEEBib1}

\end{document}

%% file: macros/vmr-symbols-vecbold.tex
\usepackage{amssymb}
\usepackage{amsfonts}
\usepackage{mathrsfs}
\usepackage{xspace}
\usepackage{bm}
\usepackage{upgreek}

\newcommand{\safemath}[2]{\newcommand{#1}{\ensuremath{#2}\xspace}}

\safemath{\bma}{\mathbf{a}}
\safemath{\bmb}{\mathbf{b}}
\safemath{\bmc}{\mathbf{c}}
\safemath{\bmd}{\mathbf{d}}
\safemath{\bme}{\mathbf{e}}
\safemath{\bmf}{\mathbf{f}}
\safemath{\bmg}{\mathbf{g}}
\safemath{\bmh}{\mathbf{h}}
\safemath{\bmi}{\mathbf{i}}
\safemath{\bmj}{\mathbf{j}}
\safemath{\bmk}{\mathbf{k}}
\safemath{\bml}{\mathbf{l}}
\safemath{\bmm}{\mathbf{m}}
\safemath{\bmn}{\mathbf{n}}
\safemath{\bmo}{\mathbf{o}}
\safemath{\bmp}{\mathbf{p}}
\safemath{\bmq}{\mathbf{q}}
\safemath{\bmr}{\mathbf{r}}
\safemath{\bms}{\mathbf{s}}
\safemath{\bmt}{\mathbf{t}}
\safemath{\bmu}{\mathbf{u}}
\safemath{\bmv}{\mathbf{v}}
\safemath{\bmw}{\mathbf{w}}
\safemath{\bmx}{\mathbf{x}}
\safemath{\bmy}{\mathbf{y}}
\safemath{\bmz}{\mathbf{z}}
\safemath{\bmzero}{\mathbf{0}}
\safemath{\bmone}{\mathbf{1}}

\bmdefine{\biad}{a}
\bmdefine{\bibd}{b}
\bmdefine{\bicd}{c}
\bmdefine{\bidd}{d}
\bmdefine{\bied}{e}
\bmdefine{\bifd}{f}
\bmdefine{\bigd}{g}
\bmdefine{\bihd}{h}
\bmdefine{\biid}{i}
\bmdefine{\bijd}{j}
\bmdefine{\bikd}{k}
\bmdefine{\bild}{l}
\bmdefine{\bimd}{m}
\bmdefine{\bind}{n}
\bmdefine{\biod}{o}
\bmdefine{\bipd}{p}
\bmdefine{\biqd}{q}
\bmdefine{\bird}{r}
\bmdefine{\bisd}{s}
\bmdefine{\bitd}{t}
\bmdefine{\biud}{u}
\bmdefine{\bivd}{v}
\bmdefine{\biwd}{w}
\bmdefine{\bixd}{x}
\bmdefine{\biyd}{y}
\bmdefine{\bizd}{z}

\bmdefine{\bixid}{\xi}
\bmdefine{\bilambdad}{\lambda}
\bmdefine{\bimud}{\mu}
\bmdefine{\bithetad}{\theta}
\bmdefine{\biphid}{\phi}
\bmdefine{\bideltad}{\delta}

\safemath{\bmia}{\biad}
\safemath{\bmib}{\bibd}
\safemath{\bmic}{\bicd}
\safemath{\bmid}{\bidd}
\safemath{\bmie}{\bied}
\safemath{\bmif}{\bifd}
\safemath{\bmig}{\bigd}
\safemath{\bmih}{\bihd}
\safemath{\bmii}{\biid}
\safemath{\bmij}{\bijd}
\safemath{\bmik}{\bikd}
\safemath{\bmil}{\bild}
\safemath{\bmim}{\bimd}
\safemath{\bmin}{\bind}
\safemath{\bmio}{\biod}
\safemath{\bmip}{\bipd}
\safemath{\bmiq}{\biqd}
\safemath{\bmir}{\bird}
\safemath{\bmis}{\bisd}
\safemath{\bmit}{\bitd}
\safemath{\bmiu}{\biud}
\safemath{\bmiv}{\bivd}
\safemath{\bmiw}{\biwd}
\safemath{\bmix}{\bixd}
\safemath{\bmiy}{\biyd}
\safemath{\bmiz}{\bizd}

\safemath{\bmxi}{\bixid}
\safemath{\bmlambda}{\bilambdad}
\safemath{\bmmu}{\bimud}
\safemath{\bmtheta}{\bithetad}
\safemath{\bmphi}{\biphid}
\safemath{\bmdelta}{\bideltad}

\safemath{\bA}{\mathbf{A}}
\safemath{\bB}{\mathbf{B}}
\safemath{\bC}{\mathbf{C}}
\safemath{\bD}{\mathbf{D}}
\safemath{\bE}{\mathbf{E}}
\safemath{\bF}{\mathbf{F}}
\safemath{\bG}{\mathbf{G}}
\safemath{\bH}{\mathbf{H}}
\safemath{\bI}{\mathbf{I}}
\safemath{\bJ}{\mathbf{J}}
\safemath{\bK}{\mathbf{K}}
\safemath{\bL}{\mathbf{L}}
\safemath{\bM}{\mathbf{M}}
\safemath{\bN}{\mathbf{N}}
\safemath{\bO}{\mathbf{O}}
\safemath{\bP}{\mathbf{P}}
\safemath{\bQ}{\mathbf{Q}}
\safemath{\bR}{\mathbf{R}}
\safemath{\bS}{\mathbf{S}}
\safemath{\bT}{\mathbf{T}}
\safemath{\bU}{\mathbf{U}}
\safemath{\bV}{\mathbf{V}}
\safemath{\bW}{\mathbf{W}}
\safemath{\bX}{\mathbf{X}}
\safemath{\bY}{\mathbf{Y}}
\safemath{\bZ}{\mathbf{Z}}

\safemath{\bZero}{\mathbf{0}}
\safemath{\bOne}{\mathbf{1}}
\safemath{\bDelta}{\mathbf{\Delta}}
\safemath{\bLambda}{\mathbf{\Lambda}}
\safemath{\bPhi}{\mathbf{\Upphi}}
\safemath{\bSigma}{\mathbf{\Upsigma}}
\safemath{\bOmega}{\mathbf{\Upomega}}
\safemath{\bTheta}{\mathbf{\Uptheta}}

\bmdefine{\biAd}{A}
\bmdefine{\biBd}{B}
\bmdefine{\biCd}{C}
\bmdefine{\biDd}{D}
\bmdefine{\biEd}{E}
\bmdefine{\biFd}{F}
\bmdefine{\biGd}{G}
\bmdefine{\biHd}{H}
\bmdefine{\biId}{I}
\bmdefine{\biJd}{J}
\bmdefine{\biKd}{K}
\bmdefine{\biLd}{L}
\bmdefine{\biMd}{M}
\bmdefine{\biOd}{N}
\bmdefine{\biPd}{O}
\bmdefine{\biQd}{P}
\bmdefine{\biRd}{R}
\bmdefine{\biSd}{S}
\bmdefine{\biTd}{T}
\bmdefine{\biUd}{U}
\bmdefine{\biVd}{V}
\bmdefine{\biWd}{W}
\bmdefine{\biXd}{X}
\bmdefine{\biYd}{Y}
\bmdefine{\biZd}{Z}

\bmdefine{\biDelta}{\Delta}
\bmdefine{\biLambda}{\Lambda}
\bmdefine{\biPhi}{\Phi}
\bmdefine{\biSigma}{\Sigma}
\bmdefine{\biOmega}{\Omega}
\bmdefine{\biTheta}{\Theta}

\safemath{\bimA}{\biAd}
\safemath{\bimB}{\biBd}
\safemath{\bimC}{\biCd}
\safemath{\bimD}{\biDd}
\safemath{\bimE}{\biEd}
\safemath{\bimF}{\biFd}
\safemath{\bimG}{\biGd}
\safemath{\bimH}{\biHd}
\safemath{\bimI}{\biId}
\safemath{\bimJ}{\biJd}
\safemath{\bimK}{\biKd}
\safemath{\bimL}{\biLd}
\safemath{\bimM}{\biMd}
\safemath{\bimN}{\biNd}
\safemath{\bimO}{\biOd}
\safemath{\bimP}{\biPd}
\safemath{\bimQ}{\biQd}
\safemath{\bimR}{\biRd}
\safemath{\bimS}{\biSd}
\safemath{\bimT}{\biTd}
\safemath{\bimU}{\biUd}
\safemath{\bimV}{\biVd}
\safemath{\bimW}{\biWd}
\safemath{\bimX}{\biXd}
\safemath{\bimY}{\biYd}
\safemath{\bimZ}{\biZd}

\safemath{\bimDelta}{\biDelta}
\safemath{\bimLambda}{\biLambda}
\safemath{\bimPhi}{\biPhi}
\safemath{\bimSigma}{\biSigma}
\safemath{\bimOmega}{\biOmega}
\safemath{\bimTheta}{\biTheta}

\safemath{\setA}{\mathcal{A}}
\safemath{\setB}{\mathcal{B}}
\safemath{\setC}{\mathcal{C}}
\safemath{\setD}{\mathcal{D}}
\safemath{\setE}{\mathcal{E}}
\safemath{\setF}{\mathcal{F}}
\safemath{\setG}{\mathcal{G}}
\safemath{\setH}{\mathcal{H}}
\safemath{\setI}{\mathcal{I}}
\safemath{\setJ}{\mathcal{J}}
\safemath{\setK}{\mathcal{K}}
\safemath{\setL}{\mathcal{L}}
\safemath{\setM}{\mathcal{M}}
\safemath{\setN}{\mathcal{N}}
\safemath{\setO}{\mathcal{O}}
\safemath{\setP}{\mathcal{P}}
\safemath{\setQ}{\mathcal{Q}}
\safemath{\setR}{\mathcal{R}}
\safemath{\setS}{\mathcal{S}}
\safemath{\setT}{\mathcal{T}}
\safemath{\setU}{\mathcal{U}}
\safemath{\setV}{\mathcal{V}}
\safemath{\setW}{\mathcal{W}}
\safemath{\setX}{\mathcal{X}}
\safemath{\setY}{\mathcal{Y}}
\safemath{\setZ}{\mathcal{Z}}
\safemath{\emptySet}{\varnothing}

\safemath{\colA}{\mathscr{A}}
\safemath{\colB}{\mathscr{B}}
\safemath{\colC}{\mathscr{C}}
\safemath{\colD}{\mathscr{D}}
\safemath{\colE}{\mathscr{E}}
\safemath{\colF}{\mathscr{F}}
\safemath{\colG}{\mathscr{G}}
\safemath{\colH}{\mathscr{H}}
\safemath{\colI}{\mathscr{I}}
\safemath{\colJ}{\mathscr{J}}
\safemath{\colK}{\mathscr{K}}
\safemath{\colL}{\mathscr{L}}
\safemath{\colM}{\mathscr{M}}
\safemath{\colN}{\mathscr{N}}
\safemath{\colO}{\mathscr{O}}
\safemath{\colP}{\mathscr{P}}
\safemath{\colQ}{\mathscr{Q}}
\safemath{\colR}{\mathscr{R}}
\safemath{\colS}{\mathscr{S}}
\safemath{\colT}{\mathscr{T}}
\safemath{\colU}{\mathscr{U}}
\safemath{\colV}{\mathscr{V}}
\safemath{\colW}{\mathscr{W}}
\safemath{\colX}{\mathscr{X}}
\safemath{\colY}{\mathscr{Y}}
\safemath{\colZ}{\mathscr{Z}}

\safemath{\opA}{\mathbb{A}}
\safemath{\opB}{\mathbb{B}}
\safemath{\opC}{\mathbb{C}}
\safemath{\opD}{\mathbb{D}}
\safemath{\opE}{\mathbb{E}}
\safemath{\opF}{\mathbb{F}}
\safemath{\opG}{\mathbb{G}}
\safemath{\opH}{\mathbb{H}}
\safemath{\opI}{\mathbb{I}}
\safemath{\opJ}{\mathbb{J}}
\safemath{\opK}{\mathbb{K}}
\safemath{\opL}{\mathbb{L}}
\safemath{\opM}{\mathbb{M}}
\safemath{\opN}{\mathbb{N}}
\safemath{\opO}{\mathbb{O}}
\safemath{\opP}{\mathbb{P}}
\safemath{\opQ}{\mathbb{Q}}
\safemath{\opR}{\mathbb{R}}
\safemath{\opS}{\mathbb{S}}
\safemath{\opT}{\mathbb{T}}
\safemath{\opU}{\mathbb{U}}
\safemath{\opV}{\mathbb{V}}
\safemath{\opW}{\mathbb{W}}
\safemath{\opX}{\mathbb{X}}
\safemath{\opY}{\mathbb{Y}}
\safemath{\opZ}{\mathbb{Z}}
\safemath{\opZero}{\mathbb{O}}
\safemath{\identityop}{\opI}

\safemath{\veca}{\bma}
\safemath{\vecb}{\bmb}
\safemath{\vecc}{\bmc}
\safemath{\vecd}{\bmd}
\safemath{\vece}{\bme}
\safemath{\vecf}{\bmf}
\safemath{\vecg}{\bmg}
\safemath{\vech}{\bmh}
\safemath{\veci}{\bmi}
\safemath{\vecj}{\bmj}
\safemath{\veck}{\bmk}
\safemath{\vecl}{\bml}
\safemath{\vecm}{\bmm}
\safemath{\vecn}{\bmn}
\safemath{\veco}{\bmo}
\safemath{\vecp}{\bmp}
\safemath{\vecq}{\bmq}
\safemath{\vecr}{\bmr}
\safemath{\vecs}{\bms}
\safemath{\vect}{\bmt}
\safemath{\vecu}{\bmu}
\safemath{\vecv}{\bmv}
\safemath{\vecw}{\bmw}
\safemath{\vecx}{\bmx}
\safemath{\vecy}{\bmy}
\safemath{\vecz}{\bmz}

\safemath{\veczero}{\bmzero}
\safemath{\vecone}{\bmone}
\safemath{\vecxi}{\bmxi}
\safemath{\veclambda}{\bmlambda}
\safemath{\vecmu}{\bmmu}
\safemath{\vectheta}{\bmtheta}
\safemath{\vecphi}{\bmphi}
\safemath{\vecdelta}{\bmdelta}

\safemath{\matA}{\bA}
\safemath{\matB}{\bB}
\safemath{\matC}{\bC}
\safemath{\matD}{\bD}
\safemath{\matE}{\bE}
\safemath{\matF}{\bF}
\safemath{\matG}{\bG}
\safemath{\matH}{\bH}
\safemath{\matI}{\bI}
\safemath{\matJ}{\bJ}
\safemath{\matK}{\bK}
\safemath{\matL}{\bL}
\safemath{\matM}{\bM}
\safemath{\matN}{\bN}
\safemath{\matO}{\bO}
\safemath{\matP}{\bP}
\safemath{\matQ}{\bQ}
\safemath{\matR}{\bR}
\safemath{\matS}{\bS}
\safemath{\matT}{\bT}
\safemath{\matU}{\bU}
\safemath{\matV}{\bV}
\safemath{\matW}{\bW}
\safemath{\matX}{\bX}
\safemath{\matY}{\bY}
\safemath{\matZ}{\bZ}
\safemath{\matzero}{\bmzero}

\safemath{\matDelta}{\bDelta}
\safemath{\matLambda}{\bLambda}
\safemath{\matPhi}{\bPhi}
\safemath{\matSigma}{\bSigma}
\safemath{\matOmega}{\bOmega}
\safemath{\matTheta}{\bTheta}

\safemath{\matidentity}{\matI}
\safemath{\matone}{\matO}

\safemath{\rnda}{A}
\safemath{\rndb}{B}
\safemath{\rndc}{C}
\safemath{\rndd}{D}
\safemath{\rnde}{E}
\safemath{\rndf}{F}
\safemath{\rndg}{G}
\safemath{\rndh}{H}
\safemath{\rndi}{I}
\safemath{\rndj}{J}
\safemath{\rndk}{K}
\safemath{\rndl}{L}
\safemath{\rndm}{M}
\safemath{\rndn}{N}
\safemath{\rndo}{O}
\safemath{\rndp}{P}
\safemath{\rndq}{Q}
\safemath{\rndr}{R}
\safemath{\rnds}{S}
\safemath{\rndt}{T}
\safemath{\rndu}{U}
\safemath{\rndv}{V}
\safemath{\rndw}{W}
\safemath{\rndx}{X}
\safemath{\rndy}{Y}
\safemath{\rndz}{Z}

\safemath{\rveca}{\bimA}
\safemath{\rvecb}{\bimB}
\safemath{\rvecc}{\bimC}
\safemath{\rvecd}{\bimD}
\safemath{\rvece}{\bimE}
\safemath{\rvecf}{\bimF}
\safemath{\rvecg}{\bimG}
\safemath{\rvech}{\bimH}
\safemath{\rveci}{\bimI}
\safemath{\rvecj}{\bimJ}
\safemath{\rveck}{\bimK}
\safemath{\rvecl}{\bimL}
\safemath{\rvecm}{\bimM}
\safemath{\rvecn}{\bimN}
\safemath{\rveco}{\bomO}
\safemath{\rvecp}{\bimP}
\safemath{\rvecq}{\bimQ}
\safemath{\rvecr}{\bimR}
\safemath{\rvecs}{\bimS}
\safemath{\rvect}{\bimT}
\safemath{\rvecu}{\bimU}
\safemath{\rvecv}{\bimV}
\safemath{\rvecw}{\bimW}
\safemath{\rvecx}{\bimX}
\safemath{\rvecy}{\bimY}
\safemath{\rvecz}{\bimZ}

\safemath{\rvecxi}{\bmxi}
\safemath{\rveclambda}{\bmlambda}
\safemath{\rvecmu}{\bmmu}
\safemath{\rvectheta}{\bmtheta}
\safemath{\rvecphi}{\bmphi}

\safemath{\rmatA}{\bimA}
\safemath{\rmatB}{\bimB}
\safemath{\rmatC}{\bimC}
\safemath{\rmatD}{\bimD}
\safemath{\rmatE}{\bimE}
\safemath{\rmatF}{\bimF}
\safemath{\rmatG}{\bimG}
\safemath{\rmatH}{\bimH}
\safemath{\rmatI}{\bimI}
\safemath{\rmatJ}{\bimJ}
\safemath{\rmatK}{\bimK}
\safemath{\rmatL}{\bimL}
\safemath{\rmatM}{\bimM}
\safemath{\rmatN}{\bimN}
\safemath{\rmatO}{\bimO}
\safemath{\rmatP}{\bimP}
\safemath{\rmatQ}{\bimQ}
\safemath{\rmatR}{\bimR}
\safemath{\rmatS}{\bimS}
\safemath{\rmatT}{\bimT}
\safemath{\rmatU}{\bimU}
\safemath{\rmatV}{\bimV}
\safemath{\rmatW}{\bimW}
\safemath{\rmatX}{\bimX}
\safemath{\rmatY}{\bimY}
\safemath{\rmatZ}{\bimZ}

\safemath{\rmatDelta}{\bimDelta}
\safemath{\rmatLambda}{\bimLambda}
\safemath{\rmatPhi}{\bimPhi}
\safemath{\rmatSigma}{\bimSigma}
\safemath{\rmatOmega}{\bimOmega}
\safemath{\rmatTheta}{\bimTheta}

%% file: macros/standard-macros.tex
\usepackage{amssymb}
\usepackage{amsfonts}
\usepackage{mathrsfs}
\usepackage{xspace}
\usepackage{bm}
\usepackage{fancyref}
\usepackage{textcomp}

\usepackage{multirow}
\usepackage{stmaryrd}

\newenvironment{textbmatrix}{	\setlength{\arraycolsep}{2.5pt}%
								\big[\begin{matrix}}{\end{matrix}\big]%
								\raisebox{0.08ex}{\vphantom{M}}}

\def\be{\begin{equation}}
\def\ee{\end{equation}}
\def\een{\nonumber \end{equation}}
\def\mat{\begin{bmatrix}}
\def\emat{\end{bmatrix}}
\def\btm{\begin{textbmatrix}}
\def\etm{\end{textbmatrix}}

\def\ba#1\ea{\begin{align}#1\end{align}}
\def\bas#1\eas{\begin{align*}#1\end{align*}}
\def\bs#1\es{\begin{split}#1\end{split}}
\def\bg#1\eg{\begin{gather}#1\end{gather}}
\def\bml#1\eml{\begin{multline}#1\end{multline}}
\def\bi#1\ei{\begin{itemize}#1\end{itemize}}

\safemath{\dirac}{\delta}					
\safemath{\krond}{\dirac}					

\safemath{\upto}{\uparrow}
\safemath{\downto}{\downarrow}
\safemath{\iu}{j}							
\safemath{\ev}{\lambda}						
\safemath{\hilseqspace}{l^{2}}				
\newcommand{\banachfunspace}[1]{\setL^{#1}}	
\safemath{\hilfunspace}{\banachfunspace{2}}	

\safemath{\SNR}{\textit{SNR}} 				
\safemath{\PAR}{\textit{PAR}} 				
\safemath{\No}{N_0}							
\safemath{\Es}{E_s}							
\safemath{\Eb}{E_b}							
\safemath{\EbNo}{\frac{\Eb}{\No}}
\safemath{\EsNo}{\frac{\Es}{\No}}

\DeclareMathOperator{\CHop}{\ensuremath{\opH}} 
\safemath{\tvir}{\rndh_{\CHop}}				
\safemath{\tvtf}{\rndl_{\CHop}}				
\safemath{\spf}{\rnds_{\CHop}}				
\safemath{\bff}{H_{\CHop}}					

\safemath{\ircf}{r_{h}}						
\safemath{\tftvcf}{r_{s}}					
\safemath{\tfcf}{r_{l}}						
\safemath{\bfcf}{r_{H}}						

\safemath{\tcorr}{c_h}						
\safemath{\scf}{c_{s}}						
\safemath{\tfcorr}{c_{l}}					
\safemath{\fcorr}{c_{H}}						

\safemath{\mi}{I}							
\safemath{\capacity}{C}						

\safemath{\normal}{\mathcal{N}}			
\safemath{\jpg}{\mathcal{CN}}			
\safemath{\mchain}{\leftrightarrow}		

\safemath{\dB}{\,\mathrm{dB}}
\safemath{\dBm}{\,\mathrm{dBm}}
\safemath{\Hz}{\,\mathrm{Hz}}
\safemath{\kHz}{\,\mathrm{kHz}}
\safemath{\MHz}{\,\mathrm{MHz}}
\safemath{\GHz}{\,\mathrm{GHz}}
\safemath{\s}{\,\mathrm{s}}
\safemath{\ms}{\,\mathrm{ms}}
\safemath{\mus}{\,\mathrm{\text{\textmu}s}}
\safemath{\ns}{\,\mathrm{ns}}
\safemath{\ps}{\,\mathrm{ps}}
\safemath{\meter}{\,\mathrm{m}}
\safemath{\mm}{\,\mathrm{mm}}
\safemath{\cm}{\,\mathrm{cm}}
\safemath{\m}{\,\mathrm{m}}
\safemath{\W}{\,\mathrm{W}}
\safemath{\mW}{\, \mathrm{mW}}
\safemath{\J}{\,\mathrm{J}}
\safemath{\K}{\,\mathrm{K}}
\safemath{\bit}{\,\mathrm{bit}}
\safemath{\nat}{\,\mathrm{nat}}

\safemath{\define}{\triangleq}			

\safemath{\equivalent}{\sim}
\safemath{\distas}{\sim}					
\safemath{\sdiff}{\Delta}				

\safemath{\reals}{\mathbb{R}}
\safemath{\positivereals}{\reals_{+}}
\safemath{\integers}{\mathbb{Z}}
\safemath{\posint}{\integers_{+}}
\safemath{\naturals}{\mathbb{N}}
\safemath{\posnaturals}{\naturals_{+}}
\safemath{\complexset}{\mathbb{C}}
\safemath{\rationals}{\mathbb{Q}}

\newcommand*{\fancyrefapplabelprefix}{app}		
\newcommand*{\fancyrefthmlabelprefix}{thm}		
\newcommand*{\fancyreflemlabelprefix}{lem}		
\newcommand*{\fancyrefcorlabelprefix}{cor}		
\newcommand*{\fancyrefdeflabelprefix}{def}		
\newcommand*{\fancyrefproplabelprefix}{prop}		
\newcommand*{\fancyrefexmpllabelprefix}{exmpl}
\newcommand*{\fancyrefalglabelprefix}{alg}		
\newcommand*{\fancyreftbllabelprefix}{tbl}		

\frefformat{vario}{\fancyrefseclabelprefix}{Section~#1}
\frefformat{vario}{\fancyrefthmlabelprefix}{Theorem~#1}
\frefformat{vario}{\fancyreftbllabelprefix}{Table~#1}
\frefformat{vario}{\fancyreflemlabelprefix}{Lemma~#1}
\frefformat{vario}{\fancyrefcorlabelprefix}{Corollary~#1}
\frefformat{vario}{\fancyrefdeflabelprefix}{Definition~#1}
\frefformat{vario}{\fancyreffiglabelprefix}{Fig.~#1}
\frefformat{vario}{\fancyrefapplabelprefix}{Appendix~#1}
\frefformat{vario}{\fancyrefeqlabelprefix}{(#1)}
\frefformat{vario}{\fancyrefproplabelprefix}{Proposition~#1}
\frefformat{vario}{\fancyrefexmpllabelprefix}{Example~#1}
\frefformat{vario}{\fancyrefalglabelprefix}{Algorithm~#1}

%% file: macros/defs.tex
\safemath{\dictab}{[\,\dicta\,\,\dictb\,]}

\safemath{\ysig}{\bmy}
\safemath{\ysighat}{\hat{\ysig}}
\safemath{\ysigdim}{M}
\safemath{\xsig}{\bmx}
\safemath{\xsigdim}{N}
\safemath{\nx}{n_x}
\safemath{\zsig}{\bmz}
\safemath{\zsigdim}{\ysigdim}
\safemath{\rsig}{\bmr}
\safemath{\Adict}{\bA}
\safemath{\Adicttilde}{\widetilde{\Adict}}
\safemath{\Adictdim}{\outputdim\times\xsigdim}
\safemath{\avec}{\bma}
\safemath{\avectilde}{\tilde{\avec}}
\safemath{\Bdict}{\bB}
\safemath{\Bdicttilde}{\widetilde{\Bdict}}
\safemath{\Cdict}{\bC}
\safemath{\cvec}{\bmc}
\safemath{\Ddict}{\bD}
\safemath{\Ddictdim}{\ysigdim\times\xsigdim}
\safemath{\dvec}{\bmd}
\safemath{\Ddicttilde}{\widetilde{\bD}}
\safemath{\Bonb}{\bB}
\safemath{\bvec}{\bmb}
\safemath{\Bonbdim}{\ysigdim\times\ysigdim}
\safemath{\noise}{\bmn}
\safemath{\noisedim}{\ysigim}
\safemath{\err}{\bme}
\safemath{\errdim}{\ysigdim}
\safemath{\errset}{\setE}
\safemath{\nerr}{n_e}
\safemath{\delop}{\bP_\errset}
\safemath{\delopc}{\bP_{{\errset}^c}}

\safemath{\cplxi}{\imath}
\safemath{\cplxj}{\jmath}

\safemath{\dict}{\matD}
\safemath{\inputdim}{N}		
\safemath{\outputdim}{M}		
\safemath{\sparsity}{S}	
\safemath{\inputdimA}{{N_a}}	
\safemath{\inputdimB}{{N_b}}	
\safemath{\elemA}{{n_a}}	
\safemath{\elemB}{{n_b}}	
\safemath{\resA}{\matR_a}	
\safemath{\resB}{\matR_b}	
\safemath{\subD}{\matS} 
\safemath{\subA}{\matS_a} 
\safemath{\subB}{\matS_b} 
\safemath{\dicta}{\matA} 	
\safemath{\dictb}{\matB} 	
\safemath{\hollowS}{H}
\safemath{\hollowA}{H_a}
\safemath{\hollowB}{H_b}
\safemath{\cross}{Z}
\safemath{\coh}{\mu_d}			
\safemath{\coha}{\mu_a}			
\safemath{\cohb}{\mu_b}			
\safemath{\mubs}{\nu}	
\safemath{\cohm}{\mu_m} 
\safemath{\dictset}{\setD}	
\safemath{\dictsetp}{\dictset(\coh,\coha,\cohb)}	
\safemath{\dictsetgen}{\dictset_\text{gen}}
\safemath{\dictsetgenp}{\dictsetgen(\coh)}
\safemath{\dictsetonb}{\dictset_\text{onb}}
\safemath{\dictsetonbp}{\dictsetonb(\coh)}

\safemath{\leftside}{U}
\safemath{\rightsideA}{R_a}
\safemath{\rightsideB}{R_b}

\safemath{\indexS}{\setI_S} 

\safemath{\na}{n_a}			
\safemath{\nb}{n_b}			
\safemath{\coeffa}{p_i}	
\safemath{\coeffb}{q_j}	
\safemath{\seta}{\setP}		
\safemath{\setb}{\setQ}     
\safemath{\setw}{\setW}	
\safemath{\setz}{\setZ}	
\safemath{\cola}{\veca}		
\safemath{\colb}{\vecb}		
\safemath{\cold}{\vecd}		
\safemath{\inputvec}{\vecx} 	
\safemath{\error}{\vece}	
\safemath{\noiseout}{\vecz} 	
\safemath{\inputvecel}{x}
\safemath{\inputveca}{\vecx_a}
\safemath{\inputvecb}{\vecx_b}
\safemath{\outputvec}{\vecy}	
\safemath{\lambdamin}{\lambda_{\mathrm{min}}}

\safemath{\elltwo}{\ell_2}
\safemath{\ellone}{\ell_1}
\safemath{\ellzero}{\ell_0}
\safemath{\ellinf}{\ell_\infty}
\safemath{\ellinftilde}{\ell_{\widetilde\infty}}
\safemath{\licard}{Z(\coh,\coha,\cohb)}
\safemath{\xsol}{\hat{x}}
\safemath{\xbord}{x_b}		
\safemath{\xstat}{x_s}		
\safemath{\xstatLone}{\tilde{x}_s}
\safemath{\order}{\mathcal{O}} 
\safemath{\scales}{\Theta} 
\safemath{\ones}{\mathbf{1}} 
\safemath{\zeroes}{\mathbf{0}} 
\safemath{\thlone}{\kappa(\coh,\cohb)} 
\safemath{\constoneA}{\delta} 
\safemath{\constoneB}{\epsilon} 
\safemath{\nlarge}{L}				   
\safemath{\sumlarge}{S_\nlarge}
\safemath{\maxlarger}{P_\nlarge}	   
\safemath{\Pzero}{\textrm{P0}}	
\safemath{\Pone}{\textrm{P1}}
\safemath{\vecfir}{\vecw}			 
\safemath{\vecsec}{\vecz}
\safemath{\elvecfir}{w}              
\safemath{\elvecsec}{z}				 
\safemath{\nlargefir}{n}
\safemath{\normout}{\gamma}
\safemath{\auxfun}{h}
\safemath{\supp}{\textrm{supp}}

\safemath{\indexa}{\ell}
\safemath{\indexb}{r}
\safemath{\indexc}{i}
\safemath{\indexd}{j}

\safemath{\project}{P}